\documentclass[aps,prl,reprint,superscriptaddress]{revtex4-1}

\usepackage{amssymb,amsmath,bm,natbib}
\usepackage[dvipsnames]{xcolor}
\usepackage[colorlinks = true,
            linkcolor = Maroon,
            urlcolor  = Maroon,
            citecolor = Maroon]{hyperref}
\usepackage{graphicx}

\begin{document}

\title{Fractional soft limits}

\author{Tom\'a\v{s} Brauner}
\email{tomas.brauner@uis.no}
\affiliation{Department of Mathematics and Physics, University of Stavanger, N-4036 Stavanger, Norway}

\author{Angelo Esposito}
\email{angeloesposito@ias.edu}
\affiliation{School of Natural Sciences, Institute for Advanced Study, Princeton, NJ 08540, USA}

\author{Riccardo Penco}
\email{rpenco@cmu.edu}
\affiliation{Department of Physics, Carnegie Mellon University, Pittsburgh, PA 15213, USA}

\date{\today}

\begin{abstract}
\noindent It is a common lore that the amplitude for a scattering process involving one soft Nambu--Goldstone boson should scale like an integer power of the soft momentum. We revisit this expectation by considering the $2 \to 2$ scattering of phonons in solids. We show that, depending on the helicities of the phonons involved in the scattering process, the scattering amplitude may in fact vanish like a \emph{fractional} power of the soft momentum. This is a peculiarity of the 4-point amplitude, which can be traced back to (1) the (spontaneous or explicit) breaking of Lorentz invariance, and (2) the approximately collinear kinematics arising when one of the phonons becomes soft. Our results extend to the general class of non-relativistic shift-invariant theories of a vector field.
\end{abstract}

\maketitle


\section{Introduction}

\noindent The modern scattering amplitude program has a two\-fold purpose~\cite{Kosower:2016cph}. On the one hand, it aims to provide alternative computation techniques for particle phenomenology, more efficient than the textbook Feynman diagram perturbation theory. On the other hand, more ambitiously, it strives to establish new foundations of quantum field theory, relying solely on physical observables rather than a Lagrangian with all its ambiguities.

Much of the effort toward understanding scattering amplitudes has been devoted to massless particles. These include, notably, gluons and gravitons~\cite{Henn:2014yza,*Elvang:2015rqa}, related to two of the fundamental interactions of Nature. However, there is another broad class of massless particles: the Nambu--Goldstone (NG) bosons of spontaneously broken symmetries. While these are rather rare in the physics of fundamental interactions, they are ubiquitous in ordered phases of matter across disciplines, including nuclear physics, condensed matter physics, astrophysics and cosmology.

In spite of the broad relevance of NG bosons, their scattering has so far been mostly studied within the framework of Lorentz-invariant field theory. Explorations of the larger landscape of field theories allowing violation of Lorentz invariance have only started to appear in the recent years. These include studies of scattering in magnetic systems~\cite{Hofmann:1998pp,*Gongyo:2016dzp}, relativistic theories with a chemical potential~\cite{Brauner:2017gkr}, rotationally invariant theories of cosmological fields~\cite{Grall:2020ibl,*Pajer:2020wnj,*Stefanyszyn:2020kay,*Bonifacio:2021azc}, cosmological solids~\cite{Jazayeri:2019nbi}, and incompressible fluids~\cite{Cheung:2020djz}. First attempts to map the entire landscape of Lorentz-breaking effective theories followed soon~\cite{Grall:2021xxm,Mojahed:2021sxy,Mojahed:2022nrn}.

As a rule of thumb, NG bosons interact weakly at large wavelengths. More precisely, in the ``single soft limit,'' in which the momentum of one NG boson participating in a scattering process goes to zero, the scattering amplitude should vanish. This property is known as ``Adler's zero.'' The rule has exceptions that have been known for a long time~\cite{Weinberg1996a}, although their proper understanding has only started to emerge recently~\cite{Kampf2020a,*Cheung2021}.

The behavior of scattering amplitudes in the soft limit can be characterized by an exponent, $\sigma$, given by the leading power-law dependence of the amplitude on the soft momentum. It is a common lore that this exponent is always a non-negative integer. The generic Adler's zero corresponds to $\sigma=1$, whereas its occasional violation to $\sigma=0$. There are also physical systems, both relativistic~\cite{Cheung2015a,Cheung2016a,Cheung2017a,*Roest:2019oiw} and non-relativistic~\cite{Mojahed:2022nrn}, where the scattering amplitudes are enhanced in the soft limit, that is $\sigma>1$. In this Letter, we show that the scaling exponent controlling the soft limit of $2 \to 2$ interactions between NG bosons may be \emph{fractional}. 

Our study was motivated by the physics of solid insulators, whose low-energy or low-temperature properties are dominated by the vibrations of the crystal lattice: the phonons. These are NG bosons associated with the spontaneous breaking of translation invariance, and their interactions are thus dictated by symmetry~\cite{Leutwyler:1996er}. In the last few years, it has become clear that a proper account of phonon--phonon interactions, and especially of their $2\to2$ scattering, is essential for understanding properties of solids such as thermal transport~\cite{Feng2017,*Gu2019,*Ravichandran2020,*Qian2021}.

We analyze in detail the 4-particle scattering amplitude of phonons in solids. We find, in accord with the common lore, that the presence of 3-phonon interaction vertices may spoil the soft limit of the amplitude altogether, leading to $\sigma=0$. However, the Adler's zero is restored if all the three non-soft (hard) phonons in the process have the same phase velocity, that is all three are either longitudinal or transverse. In this case, the resulting 3-particle kinematics is collinear, which further restricts the 4-particle scattering amplitude. When all three hard phonons are longitudinal, the ordinary Adler's zero with $\sigma=1$ is thus restored. When all three are transverse, the leading scaling exponent of the amplitude is either $\sigma=1/2$ or $\sigma=1$, depending on the precise combination of phonon helicities.

As we explain in the discussion at the end of the Letter, this peculiar behavior is not limited to phonons in solids. In fact, fractional scaling of the 4-particle amplitude is also found in a generic effective theory of a derivatively coupled 3-vector field.

Our results highlight new soft behavior of scattering amplitudes of NG bosons, specifically those with a small number of legs, which control the majority of low-energy phenomena.


\section{4-particle kinematics in the soft limit}

\noindent Consider a 4-particle scattering process involving gapless particles with linear dispersion relations, $E_i = c_i p_i$, where $p_i$ is the magnitude of momentum $\bm{p}_i$. We will use a sign convention for the momenta and energies corresponding to the scattering $1+2\to3+4$. Conservation of spatial momentum can always be used to eliminate $\bm{p}_4$ from the amplitude in favor of $\bm{p}_1$, $\bm{p}_2$ and $\bm{p}_3$. Furthermore, conservation of energy implies a relation among the relative angles of the remaining momenta,
\begin{align}
    \begin{split} \label{eq:angle23}
        \bm{p}_2\cdot\bm{p}_3 ={}& \bm{p}_1\cdot\bm{p}_2-\bm{p}_1\cdot\bm{p}_3 - \frac{c_1 c_2}{c_4^2} p_1 p_2 + \frac{c_1 c_3}{c_4^2} p_1 p_3 \\
        &+ \frac{c_2 c_3}{c_4^2} p_2 p_3 + \frac12\sum_{i=1}^3 \left(1-\frac{c_i^2}{c_4^2}\right) p_i^2\, .
    \end{split}
\end{align}
This relation can be used to remove the relative angle between $\bm{p}_2$ and $\bm{p}_3$ from the amplitude. 

Consider now the soft limit where, say, $p_1$ becomes much smaller than the magnitudes of the other momenta. It is natural to implement such a limit by rescaling the magnitude of $\bm{p}_1$ while keeping its direction fixed. Then, for a process where the speeds of propagation $c_2,c_3,c_4$ of the hard particles are all the same, the kinematics becomes degenerate. As we will see, it is this degeneracy that gives rise to the fractional soft limit in some non-relativistic theories, i.e.~theories in which Lorentz invariance is broken either spontaneously or explicitly.

While sending $p_1\to0$, it is important that on-shell energy and momentum conservation remains satisfied. In particular, when all the hard particles have the same speed,  Eq.~\eqref{eq:angle23} reduces to $\bm{p}_2\cdot\bm{p}_3 \equiv p_2 p_3 \cos \theta_{23} = p_2 p_3 + \mathcal{O}(p_1)$: the incoming particle `2' becomes approximately collinear to the outgoing particle `3' (as well as the outgoing particle `4', by conservation of momentum). This means that $\theta_{23}= \mathcal{O}\big(\sqrt{p_1}\big)$ or, equivalently, that deviations from the exact collinear limit can be parametrized as $\hat{\bm{p}}_3=\hat{\bm{p}}_2+\delta\hat{\bm{p}}_3$, with $|\delta \hat{\bm{p}}_3|=\mathcal{O}\big(\sqrt{p_1}\big)$.

We should emphasize that all these considerations apply also to the 4-particle kinematics of a relativistic theory of massless particles. In this case, all phase velocities are equal to the speed of light, and Eq.~\eqref{eq:angle23} reduces to the familiar relation among Mandelstam variables, $s+t+u=0$, which can be used to eliminate, say, $u$ from the amplitude. The remaining variables read,
\begin{subequations}
    \begin{align}
        s &= 2p_1p_2\big(1-\hat{\bm{p}}_1\cdot\hat{\bm{p}}_2\big)\,, \\
        t &= -2p_1p_3\big(1-\hat{\bm{p}}_1\cdot\hat{\bm{p}}_2-\hat{\bm{p}}_1\cdot\delta\hat{\bm{p}}_3\big)\,.
    \end{align}
\end{subequations}
Relativistic 4-point amplitudes will then also feature fractional scaling with momentum, but only at subleading orders in $p_1$. This can easily be seen, for example, in the case of four massless scalars, where Lorentz invariance forces the amplitude to be a function of $s$, $t$ and $u$ only.

When Lorentz invariance is broken, and particles with spin possibly different from zero are involved, the 4-point amplitude can have a more general dependence on the rotationally-invariant quantities built out of the particles' momenta and polarizations. As a result, the fractional powers of $p_1$ can appear in the leading contribution to the 4-point amplitude in the soft limit, as we show below.


\section{Phonons in solids}

\noindent In order to provide a concrete, phenomenologically relevant example of 4-point amplitudes that display a fractional soft limit, we are going to consider the effective theory of phonons in a homogeneous and isotropic solid. Phonon excitations are described by a 3-vector field, $\bm{\pi} (t, \bm{r})$, that parametrizes the displacement of each volume element from its equilibrium position. Homogeneity implies that a uniform translation of all volume elements should leave the system invariant. As a result, the theory must be shift invariant, and only derivatives of the phonon field can enter the effective Lagrangian. At the lowest order in the derivative expansion and up to cubic order in the fields, the Lagrangian reads $\mathcal{L} =  \mathcal{L}_2 +  \mathcal{L}_3 + \dotsb$, with
\begin{subequations}\label{eq:L}
    \begin{align} 
            \mathcal{L}_2 ={}& \bar{w} \left(\tfrac{1}{2} \dot{\bm{\pi}}^2 -  \tfrac{1}{2} \left(c_L^2 - c_T^2\right) [\Pi]^2 -  \tfrac{1}{2}c_T^2 [\Pi^T \Pi]\right) \,, \label{eq:L2} \\ 
            \begin{split}
            \mathcal{L}_3 ={}& a_1 [\Pi]^3 + a_2 [\Pi] [\Pi^2] + a_3 [\Pi] [\Pi^T \Pi] \\
            & + a_4 [\Pi^T \Pi^2] + a_5 [\Pi] \dot{\bm{\pi}}^2 + a_6 \dot{\bm{\pi}} \cdot \Pi \cdot \dot{\bm{\pi}} \,,
            \end{split} \label{eq:L3}
    \end{align}
\end{subequations}
where we have defined the matrix $\Pi_{ij} \equiv \partial_i \pi_j$, $\Pi^T$ is the transpose of $\Pi$, and $[X]$ stands for the trace of the matrix $X$. The parameters $c_L$ and $c_T$ denote the phase velocities of longitudinal and transverse phonons, while $\bar w$ is a scale controlling the dimension of the displacement field, which depends on the equilibrium properties of the system. Note that Eq.~\eqref{eq:L3} does not contain any $[\Pi^3]$ operator, since this can be eliminated in favor of $[\Pi]^3$ and $[\Pi][\Pi^2]$ using the fact that $\det \Pi$ is a total derivative (see Appendix~A of \cite{Endlich:2010hf}). Quartic interactions are not displayed in Eq.~\eqref{eq:L} since they start contributing to the 4-point amplitude only at order $\mathcal{O}\big(p_1\big)$ in the soft limit. While the quartic Lagrangian does not play a role in the following discussion, we report it for the sake of completeness in the attached Supplementary Material.

The Lagrangian~\eqref{eq:L} represents the most general (leading) kinetic term and cubic interaction that respects invariance under internal shifts, spacetime translations and spatial rotations. However, actual solids feature additional symmetries that are realized non-linearly due to being spontaneously broken, in particular Lorentz (or Galilei) boosts. Such symmetries are not manifest in the Lagrangian~\cite{Leutwyler:1996er}, but lead to relations among the coefficients of different operators. For instance, in a relativistic solid, a comparison between the quadratic Lagrangian and the stress--energy tensor of the theory shows that $\bar w$ is the relativistic enthalpy density, that is, energy density plus pressure~\cite{Esposito:2018sdc}. Moreover, one has a relation among the cubic couplings, as well as two relations between the cubic couplings and the quadratic ones. Specifically,
\begin{subequations} \label{eq:constraints}
    \begin{align} 
&a_2 - a_3 - \tfrac{1}{2}a_4 - a_5 - \tfrac{1}{2}a_6 = 0\,, \\ 
        &a_5 = \tfrac{1}{2}\bar{w}\left(1+c_L^2-2c_T^2\right)\,, \\
        &a_6 = \bar{w}\left(c_T^2-1\right)\,,
    \end{align}
\end{subequations}
with the convention that the speed of light is set to one. Such constraints are automatically implemented when the effective Lagrangian is built in a manifestly invariant way as, for example, using the coset construction~\cite{Nicolis:2013lma}. 


\section{Soft limit of the 4-phonon amplitude}

\begin{figure*}[t]
    \centering
    \includegraphics[width=0.8\textwidth]{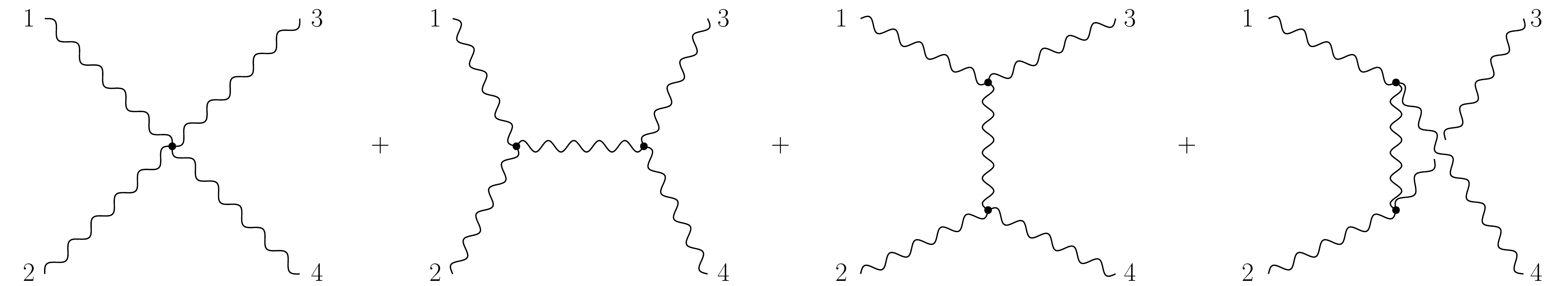}
    \caption{Feynman diagrams contributing to the on-shell tree-level 4-phonon amplitude. Each line, either an external leg or an internal propagator, can stand for a longitudinal or a transverse phonon.}
    \label{fig:feyn}
\end{figure*}

\noindent Given the Lagrangian in Eq.~\eqref{eq:L}, the amplitude for the on-shell scattering of two particles is obtained from the Feynman diagrams in Figure~\ref{fig:feyn}. The structure of the non-linear interactions makes the full amplitude too cumbersome to be displayed here. Nonetheless, it can be treated analytically, and all our results have also been checked numerically~\footnote{The expression for the 4-point amplitude can be provided in a \textsc{Wolfram Mathematica}\textsuperscript{\textregistered} format upon request.}. We also note that the quadratic Lagrangian~\eqref{eq:L2} is not canonically normalized and, therefore, each external phonon leg comes with an associated factor of $1/\sqrt{\bar{w}}$.

The soft limit of the 4-point amplitude is dominated by contributions in which the intermediate propagators of the last three diagrams in Figure~\ref{fig:feyn} go on-shell. Such contributions are related to the on-shell 3-point amplitude by a factorization formula, as dictated by standard polology arguments~\cite{weinberg_1995}. For the $s$-, $t$- and $u$-channels, we thus find respectively
\begin{subequations} \label{eq:factorization}
    \begin{align}
        \mathcal{A}^{(4)}_s &\xrightarrow{p_1\to0}-\frac{\sum_{h_\eta} \delta_{|h_\eta|, |h_2|} \mathcal{A}^{(3)}_{12\to \eta} \mathcal{A}^{(3)}_{\eta\to34}}{{(E_1+E_2)}^2-c_2^2{(\bm{p}_1+\bm{p}_2)}^2}\,, \\
        \mathcal{A}^{(4)}_t &\xrightarrow{p_1\to0}-\frac{\sum_{h_\eta} \delta_{|h_\eta|, |h_3|} \mathcal{A}^{(3)}_{1\to 3\eta} \mathcal{A}^{(3)}_{2\eta\to4}}{{(E_1-E_3)}^2-c_3^2{(\bm{p}_1-\bm{p}_3)}^2}\,, \\
         \mathcal{A}^{(4)}_u &\xrightarrow{p_1\to0}-\frac{\sum_{h_\eta} \delta_{|h_\eta|, |h_4|} \mathcal{A}^{(3)}_{1\to 4\eta} \mathcal{A}^{(3)}_{2\eta\to3}}{{(E_1-E_4)}^2-c_4^2{(\bm{p}_1-\bm{p}_4)}^2}\,,
    \end{align}
\end{subequations}
where $h_i=0,\pm1$ is the helicity of the $i$-th phonon, and the $\eta$ phonon carries the energy and momentum that flow through the propagator. The Kronecker $\delta$-functions in the numerators on the right-hand side of~\eqref{eq:factorization} ensure that only the leading part of the entire 4-point amplitude, for which the propagators go on-shell in the soft limit, is retained.

The common lore is that the presence of cubic interactions violates the Adler's zero, leading to a scaling with $\sigma=0$ in the soft limit. Indeed, in Eq.~\eqref{eq:factorization}, the 3-point amplitudes involving phonon `1' in the initial state vanish linearly with $p_1$, hence compensating for the pole of the propagator. The remaining 3-point amplitudes will, in general, be finite and non-zero in the soft limit, leading to a finite non-zero limit of the full 4-point amplitude. 

Nonetheless, there are configurations of the external phonon polarizations for which this is not true: the Adler's zero is restored. Specifically, when the speeds of the hard particles are all the same, the 4-point amplitude does vanish in the soft limit. The operational way in which the Adler's zero is restored as well as the scaling exponent with which the amplitude vanishes depends on the polarizations of the hard phonons, but is independent of the polarization of the soft one~\footnote{For longitudinal hard phonons, the contributions to the amplitude from the different channels are non-zero in the soft limit, but vanish when added together. For transverse hard phonons, instead, each channel vanishes separately. This is because on-shell 3-point amplitudes involving transverse phonons only are identically zero. This is reminiscent of the angular momentum obstruction that suppresses the collinear singularities of gluon amplitudes in quantum chromodynamics~\cite{Dixon:1996wi}. In that case, the impossibility of conserving helicity makes the collinear divergence milder. In our case, instead, the phonon's derivative couplings ensure that the amplitude is at most finite in the soft limit, while conservation of helicity makes it actually vanish. We are grateful to C.~Cheung for pointing this out.}. While the leading, $\mathcal{O}\big(p_1^0\big)$ contribution to the 4-point amplitude vanishes, the next term of the asymptotic expansion of the amplitude scales as $\mathcal{O}\big(\sqrt{p_1}\big)$. For relativistic solids where the constraints~\eqref{eq:constraints} are satisfied, its general form is
\begin{widetext}
    \begin{equation}
    \label{eq:main}
        \begin{split}
            \mathcal{A}^{(4)}_{12\to34} ={}& -\frac{\left(a_4+c_T^2 a_6\right)^2}{2\bar{w}^3 c_{T}\left(c_{T}\hat{\bm{p}}_1\cdot\hat{\bm{p}}_2-c_1\right)}(\hat{\bm{\epsilon}}_h^*\cdot\hat{\bm{\epsilon}}_1)(\hat{\bm{\epsilon}}_h^*\cdot\hat{\bm{p}}_1)(\hat{\bm{\epsilon}}_h\cdot\delta\hat{\bm{p}}_3) p_2 p_3 \\
            &\times h_2 h_3\big[\delta_{h_2,h_4}p_2+\delta_{-h_3,h_4}p_3\big]\big[\big(1+\delta_{-h_2,h_4}\big)p_2-\big(1+\delta_{h_3,h_4}\big)p_3\big]+\mathcal{O}\big(p_1\big)\,.
            \end{split}
    \end{equation}
\end{widetext} 
Recall that $p_i$ are the magnitudes of the spatial momenta. Furthermore, $\hat{\bm{\epsilon}}_1$ is the polarization vector of the soft phonon, and $\hat{\bm{\epsilon}}_h$ is the polarization vector relative to $\hat{\bm{p}}_2$ (which in the collinear soft limit equals $\hat{\bm{p}}_3$) and with helicity $h\equiv h_2h_3h_4$---i.e.~the product of the helicities of the three hard phonons. As anticipated, the source of the leading fractional soft limit is the small deviation from the collinear limit, $\delta\hat{\bm{p}}_3$, which, as mentioned above, is such that $|\delta \hat{\bm{p}}_3|=\mathcal{O}\big(\sqrt{p_1}\big)$~\footnote{Note also that, since $\bm{p}_2$, $\bm{p}_3$ and $\bm{p}_4$ are all collinear in the soft limit, the different propagators appearing in Eqs.~\eqref{eq:factorization} eventually reduce to the same structure, the one appearing in Eq.~\eqref{eq:main}.}.

We note, however, that there are helicity configurations for which the leading term~\eqref{eq:main} vanishes, thus restoring the standard linear scaling, $\sigma=1$. This happens either when the hard phonons are longitudinal, $h_2=h_3=h_4=0$, or when they are transverse with helicities such that $h_2\neq h_3=h_4$. For a correct account of the $\mathcal{O}\big(p_1\big)$ contribution to the amplitude, the Feynman diagrams with cubic interaction vertices are no longer sufficient, since they are not the only contribution at that order. One would then also need to include the contact 4-phonon diagram in Figure~\ref{fig:feyn}, arising from the quartic interaction of phonons given in the Supplementary Material. Indeed, the contribution from this diagram starts precisely at $\mathcal{O}\big(p_1\big)$.


\section{Discussion}

In this Letter we have shown, for the first time, that scattering amplitudes of NG bosons may scale with a fractional power of momentum in the soft limit. We demonstrated this explicitly using a low-energy effective theory of solids. The same conclusion, however, applies to any non-relativistic shift-invariant theory of a vector NG field, whose cubic interactions are given by Eq.~\eqref{eq:L} without further constraints on the couplings $a_i$.

The fractional scaling appears in 4-particle amplitudes where the three hard phonons are all transverse, as a consequence of the collinear kinematics of the hard phonons in the soft limit. Apart from the constraint on the phonon helicities, no particular choice of the effective couplings or phonon phase velocities is required. It is however clear from Eq.~\eqref{eq:main} that the leading, $\mathcal{O}\big(\sqrt{p_1}\big)$ contribution to such a 4-point amplitude may be canceled by a suitable choice of the $a_i$ coefficients. It might be interesting to investigate whether imposing scaling with some higher power of momentum than $1/2$ might single out a particular theory of a vector field with additional symmetry, in the spirit of~\cite{Cheung2015a,Hinterbichler:2015pqa,Cheung2018c,*Kampf:2021bet}.

Let us stress that the necessity of this peculiar fractional power scaling is limited to the 4-point amplitude. The reason is again kinematical: for higher-point amplitudes, taking the soft limit for one of the particles does not impose any particularly stringent constraint on the kinematics of the hard particles. Thus, the fractional scaling does not constitute a soft theorem that would be valid for all $n$-point amplitudes. We have checked numerically that, due to the presence of the cubic interaction vertices in Eq.~\eqref{eq:L}, a generic higher-point amplitude converges to a non-zero constant in the soft limit, i.e.~scales with $\sigma=0$.

One might also wonder what is the role played by the higher derivative corrections to the effective field theory. Because of these corrections, the phonon's dispersion relation is not exactly linear, but it is schematically given by $E(p)=c_s p\left(1+\gamma p^2 + \dotsb\right)$, where $c_s$ is the sound speed (longitudinal or transverse) and $\gamma$ is a short distance coefficient of the order of the lattice spacing squared. Within the regime of applicability of the effective theory these corrections are generally small, $\gamma p^2\ll 1$. By looking at the analog of Eq.~\eqref{eq:angle23} in the case of non-linear dispersion relations when the hard phonons have the same sound speed, it is easy to convince oneself that our considerations apply unchanged in the regime $\gamma p^2 \ll p_1/p \ll 1$, where $p$ stands for any of the three hard momenta in the $2\to2$ process. However, for $p_1/p\lesssim \gamma p^2$ the fractional scaling behavior breaks down and two things can happen: (1) for $\gamma<0$ the \emph{strict} $p_1=0$ limit is kinematically forbidden, or (2) for $\gamma>0$ the hard phonons are not collinear, but rather separated by a small angle of order $\sqrt{\gamma p^2}$. In this second instance the amplitude tends to a non-zero value at $p_1=0$.

The interesting aspect is that, in both cases, the lower cutoff $\gamma p^3$ on $p_1$, below which the qualitative behavior of the amplitude changes, is determined by an ultraviolet quantity. We thus stress how, for real solids, this is an intriguing way of estimating a short distance parameter (the lattice spacing) from a measurement in the far infrared. This is contrary to what usually happens in effective field theories, where short distance parameters are probed by shorter and shorter wavelengths.

The present work should be understood as a proof of concept. In the future, it would be interesting to pin down the precise necessary and sufficient conditions for the scattering amplitudes to display a fractional soft limit. For example, the latter could be observed also in generic $n$-point amplitudes evaluated at special kinematical points. In particular, we expect it when multiple legs are taken to be collinear, specifically $n-4$ of them, which effectively reduces the kinematics to that of a $2\to2$ scattering in the soft limit. What is peculiar to the 4-point amplitude is that the collinearity of the hard phonons is not a matter of choice, but rather is enforced by energy and momentum conservation in the soft limit.

In the same spirit, one might ask what the fate of the fractional scaling is when the momenta are promoted to complex vectors, a technique that has proven useful in different contexts (see~\cite{Brandhuber:2022qbk} for a recent review). It is clear that, for a generic set of complex momenta, one needs to give up the concept of collinearity altogether. Nonetheless, it is not immediately obvious that this will spoil the fractional scaling, at least for specific complexifications, as the ones used to probe the soft limit of NG bosons in effective field theories~\cite{Cheung2016a,Mojahed:2022nrn}.

It would also be desirable to get deeper insight into the difference in behavior of amplitudes with longitudinal and transverse phonons. Working out in detail the Ward identities for the spontaneously broken symmetries of phonons would be a good starting point. Last but not least, it would be interesting to study the soft limit of scattering amplitudes in compressible fluids, which feature an infinite number of non-linearly realized symmetries.
We leave all these directions for future work.


\section{Acknowledgments}

\begin{acknowledgments}
We are grateful to N.~Arkani-Hamed, C.~Cheung, D.~Maz\'a\v{c}, M.~Mojahed and A.~Sfondrini for useful conversations. The work of T.B.~was supported in part by the grant No.~PR-10614 within the ToppForsk-UiS program of the University of Stavanger and the University Fund. A.E.~is a Roger Dashen Member at the Institute for Advanced Study, whose work is also supported by the U.S.~Department of Energy, Office of Science, Office of High Energy Physics under Award No. DE-SC0009988. The work of R.P.~was supported in part by the National Science Foundation under Grant No.~PHY-1915611.
\end{acknowledgments}


\bibliography{biblio.bib}

\onecolumngrid

\section{Supplementary Material}
    
\noindent For completeness, we report here the quartic self-interactions of phonons in an isotropic and homogeneous relativistic solid:
\begin{align}
    \begin{split}
        \mathcal{L}_4 ={}& \tfrac{1}{4}\bar{w} \left(1-c_T^2\right) \left(\left[\Pi ^T\Pi \, \Pi ^T\Pi\right]-2 \dot{\bm{\pi}} \cdot\Pi^T\Pi \cdot \dot{\bm{\pi}}+ \dot{\bm{\pi}}^4\right)-\tfrac{1}{8}\bar{w} \left(1+c_L^2 - 2 c_T^2\right) \left(\dot{\bm{\pi}}^2-\left[\Pi ^T\Pi   \right]\right)^2 \\
        &  +\tfrac{1}{2}\left(3a_1+a_2-a_3-a_5\right)[\Pi ]^2 \left(\left[\Pi ^T\Pi  \right]-\dot{\bm{\pi}}^2\right) \\
        &+ \left(a_3+a_5\right) \left\{ 2[\Pi ]\left(\left[\Pi^T \Pi^2\right]- \dot{\bm{\pi}} \cdot \Pi\cdot \dot{\bm{\pi}} \right) - \tfrac{1}{2}  \left(\dot{\bm{\pi}}^2 - \left[\Pi^T\Pi \right]\right)\left(\left[\Pi^T\Pi\right]+\left[\Pi^2\right]\right) \right\} \\
        &+ \left(a_2-a_3-a_5\right) \left(\left[\Pi^2 \Pi^T\Pi^T\right]+\left[\Pi^T\Pi \,\Pi^T\Pi \right]+2\left[\Pi^T\Pi^3\right]-2 \dot{\bm{\pi}} \cdot \Pi^2 \cdot \dot{\bm{\pi}} -\dot{\bm{\pi}} \cdot \Pi^T\Pi \cdot \dot{\bm{\pi}} - \dot{\bm{\pi}} \cdot \Pi\,\Pi^T \cdot \dot{\bm{\pi}}\right) \\
        &  + b_1 [\Pi]^4 + b_2 [\Pi ]^2 \left(\left[\Pi ^T\Pi \right]+\left[\Pi^2\right]\right) + b_3 \left(\left[\Pi^T\Pi \right]+\left[\Pi^2\right]\right)^2 + b_4 [\Pi ] \left(3   \left[\Pi^T\Pi^2\right]+\left[\Pi^3\right]\right)\,,
    \end{split}
\end{align}
where the $b_i$ coefficients appear for the first time at this order and are independent of each other.
This expression can be derived, for instance, from the all-order Lagrangian given in~\cite{Nicolis:2013lma}.

\end{document}